# 3D printed acoustically programmable soft microactuators

**Authors:**

Murat Kaynak[1†], Amit Dolev[1†], and Mahmut Selman Sakar[1*]

**Affiliations:**

[1]Institute of Mechanical Engineering, Ecole Polytechnique Fédérale de Lausanne, CH-1015 Lausanne, Switzerland

[†] These authors equally contributed to this work.

**\*Corresponding author:**

Mahmut Selman Sakar, PhD

Institute of Mechanical Engineering, Ecole Polytechnique Fédérale de Lausanne, MED3 2916 Station 9, CH-1015 Lausanne, Switzerland.

Tel: +41 21 693 1095

Email: selman.sakar@epfl.ch

**Abstract**

The concept of creating all-mechanical soft microrobotic systems has great potential to address outstanding challenges in biomedical applications, and introduce more sustainable and multifunctional products. To this end, magnetic fields and light have been extensively studied as potential energy sources. On the other hand, coupling the response of materials to pressure waves has been overlooked despite the abundant use of acoustics in nature and engineering solutions. Here, we show that programmed commands can be contained on 3D nanoprinted polymer systems with the introduction of selectively excited air bubbles and rationally designed compliant mechanisms. A repertoire of micromechanical systems is engineered using experimentally validated computational models that consider the effects of primary and secondary pressure fields on entrapped air bubbles and the surrounding fluid. Coupling the dynamics of bubble oscillators reveals rich acoustofluidic interactions that can be programmed in space and time. We prescribe kinematics by harnessing the forces generated through these interactions to deform structural elements, which can be remotely reconfigured on-demand with the incorporation of mechanical switches. These basic micromechanical systems will serve as the building blocks for the development of a novel class of untethered soft microrobots powered and controlled by acoustic signals.

**Introduction**

Soft materials can be programmed to change their physical properties such as shape and stiffness on-the-fly by the externally applied light, heat, mechanical forces or magnetic fields[1–5]. Such material formulations have great potential, particularly at small scales, to achieve functionalities that are unattainable by conventional mechatronic systems[6–13]. Notably, simple mechanical structures fabricated from magnetorheological or liquid crystal elastomers displayed a virtually unlimited number of degrees of freedom (DOF), as a result of either the spatial complexity of the magnetization profile[14–16] or the use of structured illumination[17,18]. Moreover, rationally designed flexible structures, also known as mechanical metamaterials, can realize programmable digital logic[19–29]. This route for physical intelligence has already been pursued by living organisms[30–32], and extensively studied for the development of autonomous soft robots[33–36].

Acoustofluidics has several unique properties that can reveal the full potential of programmable soft matter. Air bubbles are efficient transducers that generate forces through interactions with the external pressure waves, the surrounding fluid, and with each other[37–45]. Therefore, these interactions can be programmed by modulating the acoustic wave as well as the geometry and the spatial distribution of bubbles. The generated forces are determined by primary and secondary acoustic effects that are highly nonlinear. Thus, a detailed understanding of the underlying physics is instrumental to use air bubbles for controlling deformable elements. To address this unmet challenge, we combined analytical modeling, finite element simulations, and experimental analysis that involved detailed characterization of acoustic pressure, bubble oscillations, and forces generated by bubbles with coupled dynamics.

Here, we present an integrated design, fabrication, and control methodology that transforms monolithically printed flexible structures into programmable soft robotic systems. The key innovation is the spatial patterning of polymer capsules that stably contain individual air bubbles along ultraflexible beams using direct laser writing. By tuning the geometry of the capsules and the architecture of the material, we precisely controlled the acoustofluidic interactions. We demonstrate the compatibility of our acousto-active systems with mechanical logic by constructing actuated bistable mechanisms.

**Results and Discussion**

The basic actuator module consists of a 3D printed cylindrical capsule with a single circular orifice (Fig. 1a). We printed all the structures as a monolithic piece from a single biocompatible soft polymer, trimethylolpropane ethoxylate triacrylate (TPETA)[46], using two-photon polymerization (Fig. 1b). A detailed description of the experimental methods is given in Supplementary Note 1. Within this capsule, the bubble is physically isolated from the surrounding fluid except at the side of the opening (Fig. 1c). Acoustic excitation of a single bubble submerged in liquid generates microstreaming.[47] By entrapping bubbles inside cavities, their oscillations are constrained to regions defined by the orifices. In this configuration, acoustic streaming generated counter-rotating vortices and a jet localized in the center, which was captured using high-speed recordings of tracer particles around the actuator module (Fig. 1d). Once an actuator is connected to a clamped cantilever beam, it is expected to generate thrust normal to the orifice's plane, thereby deforming the beam. By carefully tuning the acoustic frequency, different vibration modes of the coupled fluid-structure system can be excited. The vibration modes are manifested by the unique deformation patterns of the bubble at the interface. Nevertheless, we focused our attention to the first mode because, assuming a uniform distribution, the projection of the impinging pressure is maximal on this mode.

Large deformation and complex motion can be generated with slender structural elements including beams, plates, and shells[48–50]. We based our design methodology on beams and flexures because they are easier to manufacture, experimentally characterize, and computationally model. We have recently developed an analytical model that calculates the natural frequencies and corresponding vibration modes of bubbles entrapped inside arbitrarily shaped cavities with multiple circular orifices[51]. In brief, we extended a previously published model[52] to accommodate multiple orifices on the same cavity, incorporate all the possible mode shapes, and consider the geometry adjacent to the office.

We recorded the power spectrum of the ultrasound transducer using a hydrophone to determine the frequency bands at which the actuators could be effectively powered (Fig. S1). Using our model, we designed the geometry (i.e., volume and orifice radii) so that the first natural frequencies of the entrapped bubbles are within these bands (Fig. 1e). Accordingly, we printed two capsules with orifice radii of 7.75 μm and 13 μm, respectively (Table S1). Entrapped bubbles were actuated selectively at their estimated natural frequencies of either 85 kHz or 170 kHz, which led to a sequential deformation of the beams (Fig. 1f and Movie S1).

Mechanisms printed parallel to the substrate are not ideal for the quantification of forces because the bubbles may interact with the substrate's surface[44,45]. To minimize such perturbations, we printed the beams vertically, raising the actuator modules 50 μm above the substrate (Fig. 2a). Laser scanning confocal microscope images of fluorescently labeled samples verified that the structures were printed according to the CAD design (Fig. S2). In this configuration, beam bending could be followed from the actuator's in-plane displacement (Fig. 2b), which increased quadratically with the input voltage (Fig. 2c). The bending modulus was calculated from the material's Young's modulus, E = 13 MPa[46], and the dimensions of the beam were measured using electron microscopy (Fig. S2). The generated force was then estimated using linear beam theory (Supplementary Note 2). Although the linear theory was used, the displacement is a nonlinear function of the applied force because the part of the beam corresponding to the location of the actuator was taken as rigid. The tip displacement, $δ$, is given as:

$$\delta = \frac{LP}{6EI}\left[ L(3l+2L) + \frac{6l(2l+L)}{\sqrt{4+\frac{L^2(2l+L)^2 P^2}{E^2 I^2}}} \right] \quad (1)$$

where $L$ is the length of the beam, $l$ is the length of half the actuator, $E$ is Young's modulus, $I$ is the second moment of inertia of the cross-section and $P$ is the applied force. Finite element method (FEM) simulations of a nonlinear model were in perfect agreement with the analytical model, thus confirming that the linearity assumption was acceptable for the given range of deflections (Fig. 2d). The analytical model was used to calculate the total force acting on a deforming beam from the recorded deflection.

We observed a linear relationship between the voltage applied to the transducer and the resultant acoustic pressure measured by a hydrophone (Fig. 2e), as predicted by the theory[53]. The acoustic energy density in the workspace quadratically increases with increasing pressure[54], and so does the streaming velocity around the bubble[44]. Sequentially increasing the input voltage resulted with a new configuration within milliseconds. Furthermore, when the input voltage was turned off the beam immediately returned to its original position, confirming that the actuation was in the elastic range. We characterized the oscillation amplitude at the bubble's first natural frequency using a custom-built experimental platform (Fig. S3). The data showed that the deflection amplitude at the air-fluid interface increases linearly for the range of pressure we applied (Fig. 2f).

An actuator module with multiple DOF would drastically increase the dexterity and form factor of printed structures. We postulated that a capsule that contains multiple orifices of different sizes would serve this purpose. To test this idea, we printed an actuator module in the shape of an equilateral triangular prism with three orifices at the center of each face (Fig. 3a). The actuator module was connected atop a 50 µm long and 2 µm diameter cantilever beam. The sizes of the orifices (15, 9.5 and 7 µm) were chosen in a way that corresponding natural frequencies were well spaced and within the effective range of the transducer (Fig.3b). Simulation results have shown that primarily one interface is deformed at each mode (Fig. 3c). We excited the system at multiple frequencies simultaneously to control the position. The force is unidirectional, therefore, at least three frequencies are required to fully control the 2D position. Ideally, these frequencies are the ones where the largest displacement is obtained, which are expected to be the bubble's natural frequencies. However, the transducer has its own dynamical response, and the input pressure varies with the excitation frequency. We addressed this issue by experimentally calibrating the response of the system.

We recorded the actuator motion while sweeping the frequency from 40 kHz to 160 kHz at a constant input voltage. Using a subpixel resolution image processing algorithm[56], we extracted the planar position of the module (Fig. S4). We then selected three frequencies where the largest deflections in different axes were recorded. Figure 3d shows the undeformed and deformed states of the system at the chosen frequencies. Next, we measured the actuator's deflection at each frequency and built the vectors spanning the planar displacement field. We implemented a controller that leverages the superposition principle, and excites the system with a signal comprising three harmonic terms which corresponds to the aforementioned vectors. We control two DOF with three forces, thus, there are infinite combinations. We solved a constrained linear least-squares problem[57], where we minimized the total amplitude of the applied signal.

We performed a series of experiments to evaluate the control strategy. First, we programmed a calibrated device to repeatedly follow a rectangular trajectory, slightly offset from the origin where we had better control over the motion (Fig. 3e). Second, we programmed several circular trajectories with different radii and offsets (Movie S2). The position error increased with the distance from the origin (Fig.

3f). Third, we used the calibration parameters of one device to control another with the same design (Fig. S4). The trajectories of the two devices were close, suggesting that calibration might be performed once and used repeatedly for different devices. As a final demonstration, we designed four trajectories prescribing the initials EPFL to show that control was not limited to specific geometries (Fig. 3g).

Acoustically excited bubbles interact with each other through the surrounding fluid when they reside in close proximity. The total force acting on a bubble is the result of the exciting primary pressure field, and higher order fields emanating from neighboring bubbles that also act as acoustic sources[38,58]. The distances between bubbles are significantly smaller than the acoustic wavelength in all our prototypes. Therefore, we can assume that the acoustic radiation forces caused by the primary field do not affect the relative displacement of the actuators. The secondary forces that act on the coupled actuators are thrust, drag-induced acoustic streaming, and secondary acoustic radiation force (also known as secondary acoustic radiation force or secondary Bjerknes force). Identical bubbles are expected to generate the same acoustic streaming; thus, the generated thrust would push the bubbles away from each other. On the other hand, the magnitude of the acoustic radiation force, which primarily acts to pull the bubbles toward each other, depends on the distance between them[38,58]. The total force, $F_B$, acting on a bubble is

$$F_B = F_R + F_{AS} + F_d \qquad (2)$$

where $F_R$ denotes the acoustic radiation force, $F_{AS}$ denotes thrust generated by streaming, and $F_d$ is the drag force acting on a bubble due to the streaming generated by the adjacent bubble.

To quantify the total force generated by interacting bubbles, we printed two adjacent cantilever beams with identical actuators that were faced toward each other (Table S1). The initial distance between the actuators, $d_i$, was systematically varied to study the effect of spacing. At equilibrium, $F_B$ in Eq.(2) represent the elastic force applied by the cantilever beam, denoted by P in Eq.(1). We observed two distinct regimes in the dynamics of the coupled actuators (Movie S3). When the initial distance between the actuators was smaller than a critical distance, $d_c$, the radiation forces dominated the thrust generated by acoustic streaming (Fig. 4a). For the given actuator design, this critical distance was 50 µm (i.e., $d_c$ = 50 µm). As a result, the beams bent towards each other until the bubbles made contact (Fig. 4b). With increasing $d_i$, the magnitude of the acoustic radiation force decreased, emphasizing the contribution of streaming forces (Fig. 4c). For $d_i > d_c$, the sign of the total force switched, where the beams started to move away from each other (Fig. 4d).

Figure 4e summarizes the nonlinear responses of coupled actuators with respect to the initial distance and the input voltage. The further away the actuators were from each other at rest, the more the beam deflection resembled that of the isolated single beam (see Fig. 2c). Based on this empirical observation, we hypothesized that the dynamics could be captured by an analytical model where all forces are proportional to the input voltage squared[55]. We assumed that $F_d$ did not depend on the distance between the actuators, therefore $F_{AS} = F_d$ at all times. This assumption is reasonable as the distance between the bubbles was always comparable to the bubble size, which was significantly shorter than the acoustic wavelength. We also assumed that $F_R$ inversely depend on the distance squared[38,58]. The force balance equation, Eq.(2), is then re-written as:

$$F_B = (\alpha d^{-2} + 2\gamma)V^2 \qquad (3)$$

where $\alpha$ and $\gamma$ are functions of the excitation frequency and geometry, and $d$ is the distance between the bubbles' vibrating surface. We fitted $\alpha$ = 174.5 nN V$^{-2}$ µm$^{-2}$, and $\gamma$ = 0.027 nN V$^{-2}$ at 125 kHz to the empirical data shown in Figure 3e. The model could capture the dynamics represented in the experimental data (Fig. S5). This analytical model has been used to design the prototypes presented in the rest of the article.

Our results have shown that radiation forces between identical actuator modules do not change direction for a given spacing. Actuators with different orifice sizes displayed more complex interactions. We discovered that depending on the excitation frequency, actuators with the same initial distance attracted or repelled each other. An analogous phenomenon was observed between acoustically excited free-floating spherical bubbles with different radii[38,58]. Bubbles with distinct natural frequencies oscillate with a relative phase that dictates the direction of the radiation force. When the bubbles oscillate in a relative phase of less than a quarter of a period, the force is attractive, but if the phase differ by more than a quarter but less than three quarters of a period, the force is repulsive (Fig. 4f). Therefore, identical bubbles tend to attract each other, and non-identical bubbles can either attract or repeal each other, depending on the excitation frequency (Fig. 4g). In our experiments, actuator modules with cavity radii of 25 µm and 17.5 µm and orifice radii of 10 µm, attracted or repelled one another at 85 kHz and 125 kHz, respectively (Fig. 4h and Movie S4).

We designed a flextensional mechanism that leverages this frequency-dependent behavior to manifest multiple distinct deformation patterns on the same system (Fig. S6 and Movie S5). Both couples of arms simultaneously closed at one frequency, and one couple opened while the other closed at another frequency. The frequencies at which the arms would open or close were determined by the relative phase of oscillations, which was modified by capsule geometry. Both operation modes were independent of the input signal amplitude, therefore, the angle between the arms could be tuned with the applied voltage.

So far, we focused on a paradigm where actuator modules were patterned on different structures. In this arrangement, rapid increase in radiation forces with decreasing distance limits the range of motion that the actuators can generate. To extend the interval at which the structure bends in a graded fashion, we constrained the actuators' motion. To this end, we connected two actuators with a truss so that they were not allowed to come very close to each other (Fig. 5a and Fig. S7). As expected, the arms progressively bent out of plane for a large range of input voltage (Movie S6). To report the deformation, we recorded the displacement of the actuators along the y-axis (Fig. 5b). Although the arms bent under the dominant radiation forces, the displacement curve did not follow the highly nonlinear trend presented in Figure 4e. We built a FEM model of the mechanism based on the CAD design and the material properties (Fig. S7). The radiation force magnitude and direction are expected to change as the arms bend due to the relative position of the bubbles. We simplified the model by assuming that the radiation force always acted to pull the actuators towards each other, and its magnitude was proportional to the thrust. We took the thrust calculated for a single actuator module (see Fig. 2) as an input and estimated the acoustic radiation force for the applied voltage values by fitting the empirical data shown in Figure 5b. Here, the acoustic radiation force was taken as $F_R = \beta V^2 d^{-2}$ following Eq.(2), where $\beta$ was estimated as 745.6 nN V$^{-2}$ µm$^{-2}$ (Supplementary Note 3). In the experiments, at relatively high voltage values, further increment did not cause further deformation (Fig. 5b). The experimentally observed plateau may be due to the drag force applied to the moving actuators by the anchored ones.

The out-of-axis bending is a classic example for unimorph actuators where we control the angular displacement. By simply connecting a microbubble pair with an ultraflexible spring mechanism, we developed a linear microactuator (Fig. 5c). Acoustic forces were primarily uniaxial and, as expected, we did not observe out-of-axis deflection during operation (Movie S7). Attractive radiation forces between the two microbubbles caused the gap to narrow. The deflection of the spring followed an almost linear trend with respect to applied voltage until the bubbles were 20 µm apart from each other, at which point the radiation forces snapped the mechanism (Fig. 5d). When the excitation signal was turned off, the acoustic forces vanished, and the actuator returned to its initial position. We calculated the magnitude of forces that correspond to the input voltage using Eq.(3) and applied these forces to an FEM model. The model captured the behavior of the experimentally recorded spring deflection, verifying that we have a reliable way of calculating forces generated by the acoustic actuator modules (Fig. 5e). We calculated the stiffness of the spring as 0.7563 nN/µm by computing the derivative of the displacement with respect to the total force (Fig. S8).

Constrained elastic beams can exhibit complex mechanical responses depending on the geometry, degree of confinement, and boundary conditions. Previous work has shown that such mechanisms instantiate embodied logic and programmable functionality in soft machines[19,33]. We designed beams to present a snap-through instability so that application of a relatively small thrust would be sufficient to cross the energy barrier and trigger rapid and large deformation towards a second stable configuration. Key geometric parameters for the beam design are the inclination angle of the beam, $\theta$, and the slenderness ratio, $w/L$, where $w$ and $L$ denote the width and length of the beam, respectively (Fig. 6a). The mechanism was driven by a single actuator module that operated in acoustic streaming mode (Fig. 6b). The modules stayed indefinitely in both stable states, and the thrust generated by the actuator module was high enough to pass the energy barrier.

For a fixed beam length and cross-sectional profile, as $\theta$ increases, the input pressure required to switch the mechanism state is expected to increase while the deformed position becomes more stable[20]. We fabricated three different prototypes that only differ in inclination angle ($\theta$ = 30°, 45°, and 60°) to validate the theoretical predictions. We observed a monotonic increase in the pressure at which the beam managed to switch states (Fig. 6c). We used a 2D FEM model to obtain the double well potential energy landscape for the same design parameters (Fig. 6d). The simulation results showed that the strain energy quadruples when $\theta$ is increased from 30° to 70°, and the module displacement doubles. We calculated the force required to switch the mechanism at different $\theta$ from the empirical data using Eq.(3), where we only considered the thrust generated by a single actuator module (i.e., $F_B = 0.027V^2$). Comparing these values with the simulated force showed that the switching occurred at lower levels than predicted (Table S2). The rationale behind this discrepancy is that the actuator module did not move along a straight line as simulated, and, instead, moved in 3D by following a more favorable energy landscape.

We harnessed frequency selective thrust generation to realize reversible actuation for the bistable mechanism (Fig. 6e). We kept the bubble size constant and tuned the actuators' orifice size to be able to activate them at distinct frequencies, 85 kHz and 170 kHz (see Fig. 1f). Connecting two actuator modules with a single beam to the anchor point proved to be undesirable. To stabilize the structure and ensure reliable operation, we extended the mechanism by adding support structures. The resulting mechanism could be switched repeatedly at the same amplitude and excitation frequency (Movie S8).

As a final demonstration, we programmed the motion of a continuously bending cantilever beam using an actuated bistable mechanism, which we refer to as the control module (Fig. 6f-left). Here, triggering the control module changes the direction of bending by introducing radiation forces to a system otherwise driven solely by acoustic streaming (Movie S9). To trigger the controller and actuate the beam at different frequencies, we engineered capsules with different geometries. At 125 kHz, the beam bent clockwise while the control module stayed idle (Fig. 6f-middle). Exciting the system at 85 kHz activated the bubble located on the left of the control module, moving the right one closer to the actuator connected to the beam. The control module stayed in this stable state when the system was turned off, as expected. Then, exciting the system again at 125 kHz bent the beam counterclockwise due to the attractive radiation forces generated between the two neighboring bubbles (Fig. 6f-right). In this prototype, the control module must be reset manually because the radiation forces are stronger than the thrust. However, by tuning the bubbles' geometry, we can introduce a third frequency at which the control module would be reset.

**Conclusions**

We introduced a suite of mechanical microsystems with relatively basic design yet complex motion. Precise control over geometric parameters facilitated detailed analysis of forces and fluid-structure interactions. We made two important discoveries. First, we showed that a single bubble can provide multiple-degrees-of-freedom motion along prescribed trajectories with high fidelity through the application of frequency and amplitude modulated acoustic signals. The realization of this feature required printing of capsules with multiple orifices at different sizes and a real-time calibration process. Second, as a result of higher order interactions, forces between oscillating bubbles emerge, and their direction can be programmed by tuning the excitation frequency.

We developed a control scheme that drives a mechanical system to follow prescribed trajectories by applying a complex signal. The position errors between the programmed and executed motion are not negligible, which may stem from multiple sources. We assumed that the 3D printed cantilever beam has an isotropic stiffness. However, the material stiffness depends on the degree of polymerization, which might show spatial variations along the vertical axis. Moreover, we assumed small deflections although the beam deflected beyond the limit defined by the theoretical linear model. Finally, at relatively large deflections, the beam started to tilt out of plane and rotate around its axis while we only considered in-plane translation.

We focused our attention on geometric arrangements of entrapped bubbles, as a means to program the soft robotic system. Leveraging frequency selective bubble excitation, we could operate systems in different modes at different frequencies. Considering the interactions among neighboring actuators would advance the programming framework. An actuator may be attracted to other actuators from multiple directions, realizing a sophisticated motion that highly depends on the excitation frequency and time-varying spacing among bubbles. We have also incorporated acoustically actuated bistable mechanisms as on-board control units to introduce the capability to operate in different modes under the same input signal. This last functionality represents a first step towards reconfigurable mechanical systems that operate analogous to mechatronic systems. To this end, reversible switching of several bistable structures will be instrumental. Moving forward, bistable beams could be replaced by 3D architected materials that display sensing, pattern analysis, and multi-stability[61–63]. The fabrication of such complex structures is feasible with two-photon lithography, and local actuation initiated by entrapped bubbles will reveal unprecedented reconfigurability and multifunctionality.

**Acknowledgements**

We thank Dr. Guillermo Villanueva for generously providing the vibrometer, Dr. Pedro Reis for fruitful discussions, and Furkan Ayhan for his assistance with electron microscopy imaging.

**Author Disclosure Statement**

Authors declare no competing financial interests.

**Funding Information**

This work was supported by the European Research Council (ERC) under the European Union's Horizon 2020 research and innovation program (Grant agreement No. 714609).

Matter. Proc. Natl. Acad. Sci. U. S. A. 2016;113(41):E6007–E6015.

15. Hu W, Lum GZ, Mastrangeli M, Sitti M. Small-Scale Soft-Bodied Robot with Multimodal Locomotion. Nature. 2018;554(7690):81–85.

16. Kim Y, Yuk H, Zhao R, Chester SA, Zhao X. Printing Ferromagnetic Domains for Untethered Fast-Transforming Soft Materials. Nature. 2018;558(7709):274–279.

17. Palagi S, Mark AG, Reigh SY, Melde K, Qiu T, Zeng H, Parmeggiani C, Martella D, Sanchez-Castillo A, Kapernaum N, *et al.* Structured Light Enables Biomimetic Swimming and Versatile Locomotion of Photoresponsive Soft Microrobots. Nat. Mater. 2016;15(6):647–653.

18. Rogóż M, Zeng H, Xuan C, Wiersma DS, Wasylczyk P. Light-Driven Soft Robot Mimics Caterpillar Locomotion in Natural Scale. Adv. Opt. Mater. 2016;4(11):1689–1694.

19. Song Y, Panas RM, Chizari S, Shaw LA, Jackson JA, Hopkins JB, Pascall AJ. Additively Manufacturable Micro-Mechanical Logic Gates. Nat. Commun. 2019;10(1):882.

20. Jiang Y, Korpas LM, Raney JR. Bifurcation-Based Embodied Logic and Autonomous Actuation. Nat. Commun. 2019;10(1):1–10.

21. Meng Z, Chen W, Mei T, Lai Y, Li Y, Chen CQ. Bistability-Based Foldable Origami Mechanical Logic Gates. Extrem. Mech. Lett. 2021;43101180.

22. Raney JR, Nadkarni N, Daraio C, Kochmann DM, Lewis JA, Bertoldi K. Stable Propagation of Mechanical Signals in Soft Media Using Stored Elastic Energy. Proc. Natl. Acad. Sci. 2016;113(35):9722–9727.

23. Bertoldi K, Vitelli V, Christensen J, Van Hecke M. Flexible Mechanical Metamaterials. Nat. Rev. Mater. 2017;2.

24. Treml B, Gillman A, Buskohl P, Vaia R. Origami Mechanologic. Proc. Natl. Acad. Sci. 2018;115(27):6916–6921.

25. Preston DJ, Rothemund P, Jiang HJ, Nemitz MP, Rawson J, Suo Z, Whitesides GM. Digital Logic for Soft Devices. Proc. Natl. Acad. Sci. U. S. A. 2019;116(16):7750–7759.

26. Kadic M, Milton GW, van Hecke M, Wegener M. 3D Metamaterials. Nat. Rev. Phys. 2019;1(3):198–210.

27. Pal A, Restrepo V, Goswami D, Martinez R V. Exploiting Mechanical Instabilities in Soft Robotics: Control, Sensing, and Actuation. Adv. Mater. 2021;33(19):2006939.

28. Yasuda H, Buskohl PR, Gillman A, Murphey TD, Stepney S, Vaia RA, Raney JR. Mechanical
16

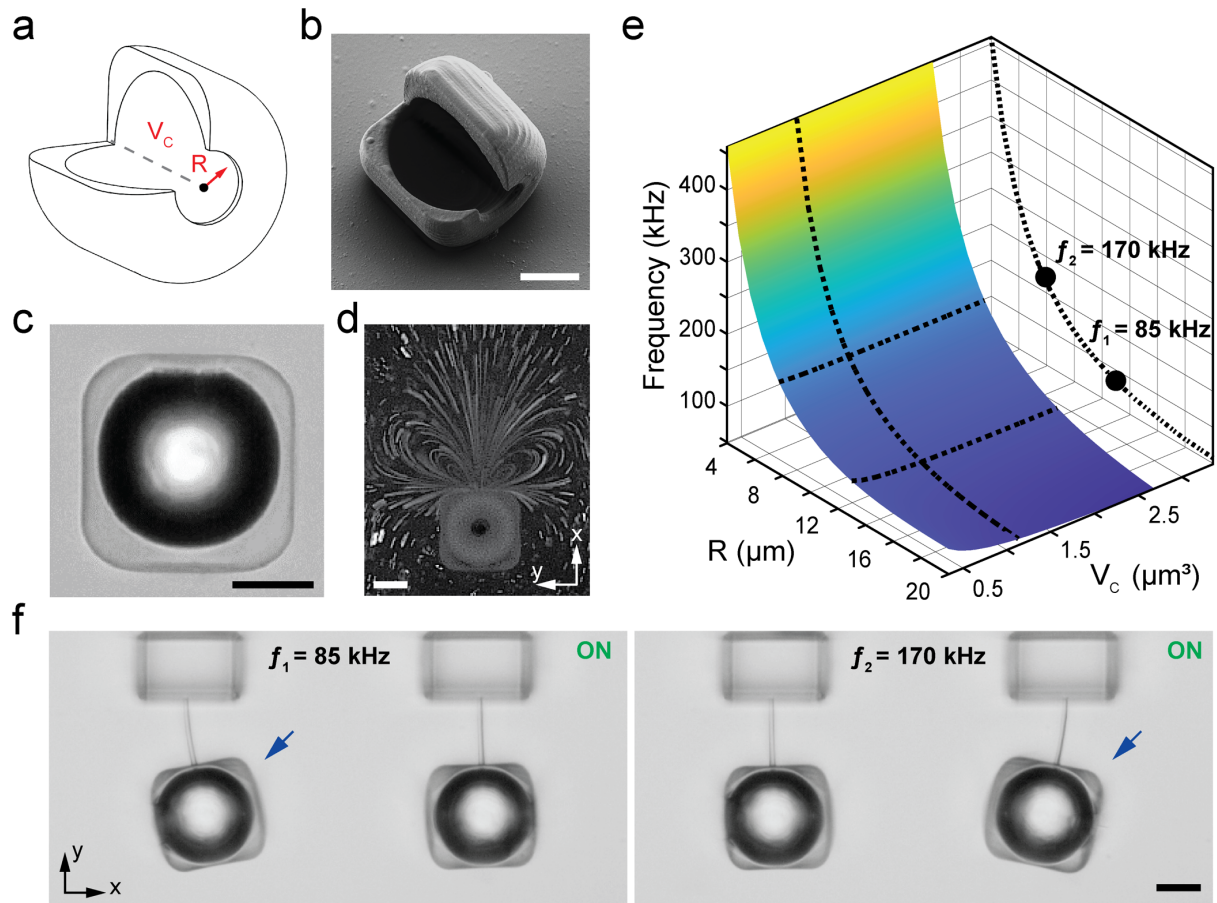

**Fig. 1. Design, fabrication and operation of acoustic actuator modules. a** The natural frequency of oscillations of entrapped microbubbles were set by tuning the volume of the cavity, $V_C$, and the radius of the orifice, R. **b** Scanning electron microscopy image of a partially printed actuator module shows the cavity and the orifice. **c** Brightfield images of an actuator module when the air bubble is entrapped inside the cavity. **d** Streamlines around an acoustic module are visualized using fluorescent tracer particles. **e** A computed surface plot showing the dependence of natural frequency of the entrapped air bubble on the orifice size (R) and cavity volume ($V_C$). The black dashed lines highlight constant values along the surface, which correspond to the selected geometric parameters. A projection of the constant volume line is depicted with two marker points highlighting the desired frequencies. **f** Frequency-selective powering of multiple actuator modules. The first natural frequencies of the two modules are set by prescribing different orifice sizes. The blue arrow points to the actuator module that is activated at the specified frequencies, $f_1$ and $f_2$. Scale bars, 20 μm.

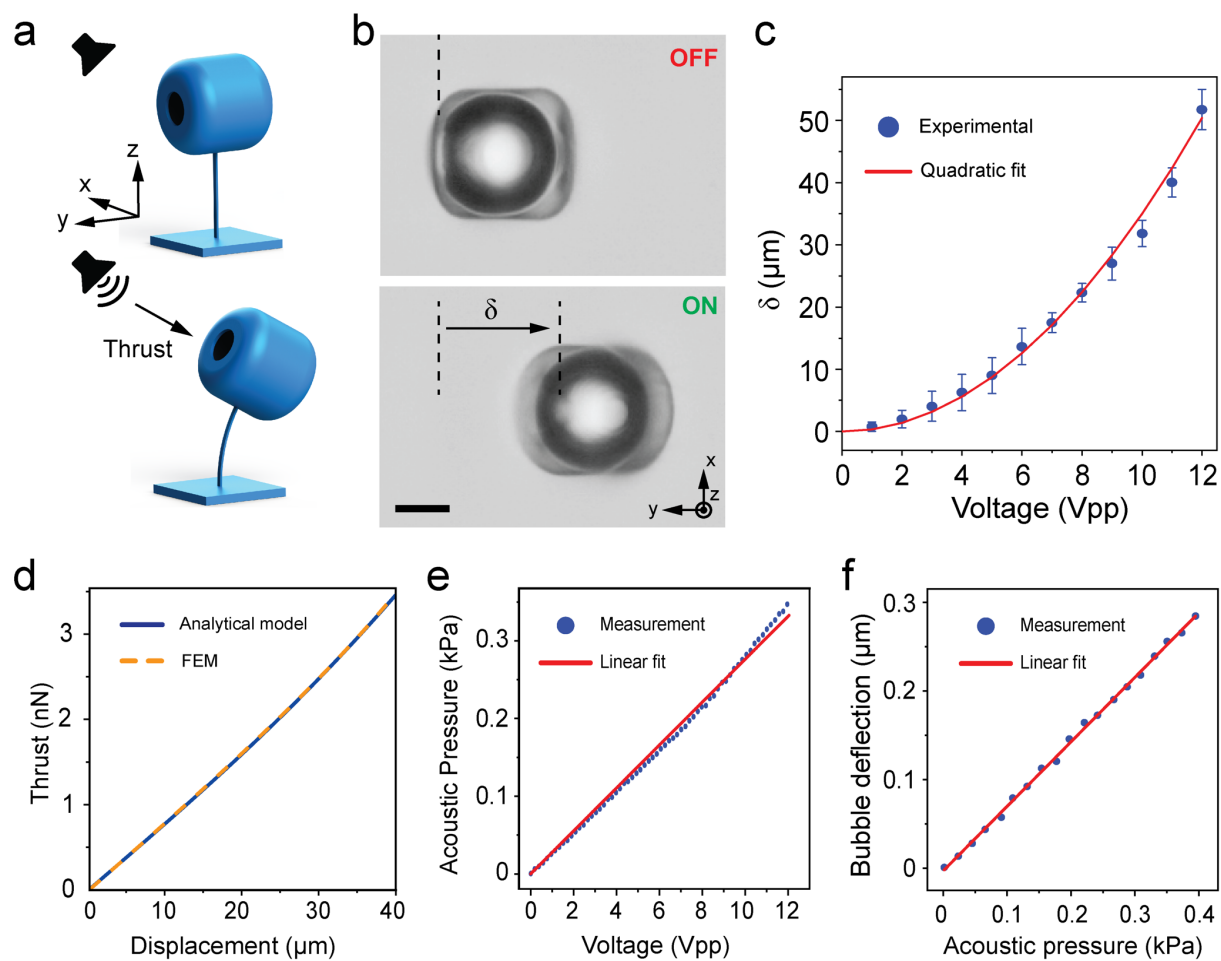

**Fig. 2. Mechanical characterization of the actuator module and the pressure field. a** Schematic illustration of the bending of a flexible cantilever beam that is printed vertically to minimize the interactions of the actuator module with the bottom surface. **b** Brightfield images showing the displacement of the actuator module, which is denoted as δ. δ is used to measure the deflection of the beam. Scale bar, 20 μm. **c** The displacement increases quadratically with the voltage applied to the acoustic transducer. **d** The thrust calculated using a linear analytical model and an FEM model are in excellent agreement. **e** Acoustic pressure increases linearly with the voltage applied to the transducer. Pressure measurements are performed using a hydrophone. **f** The deflection of the entrapped bubble measured at the center of the orifice linearly increases with the acoustic pressure. The deflection was measured using a laser vibrometer.

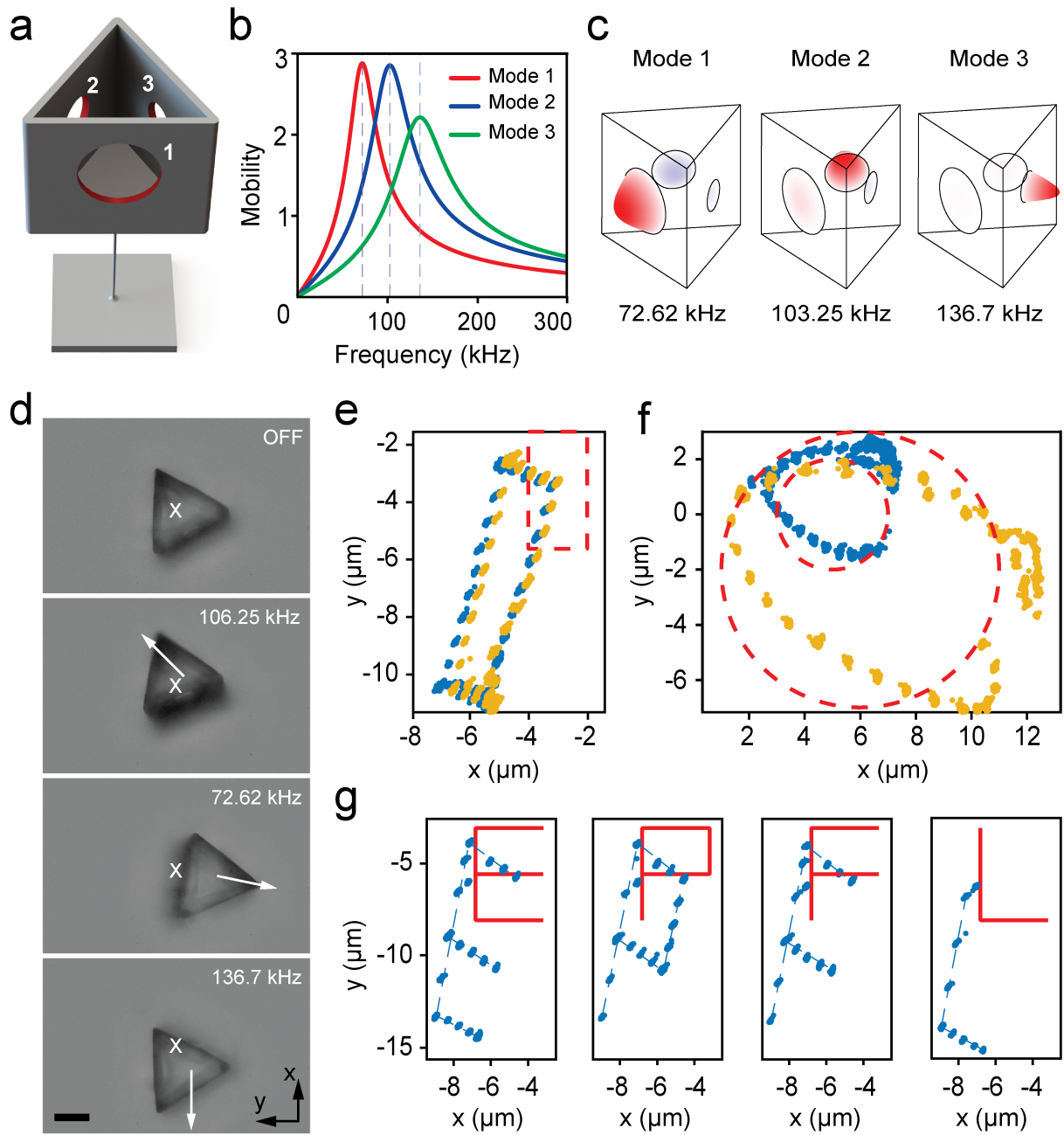

**Fig. 3. Design, calibration, and control for multi DOF motion. a** Illustration of the actuator module with multiple orifices. The ceiling is removed for visualization. The different orifices are numbered starting from the largest to the smallest. The radii are 15 μm, 9.5 μm, and 7.5 μm, respectively. **b** Modal mobility plot, showing the first three modes of the system excited with uniform pressure. Dashed lines highlight the natural frequencies. **c** Illustration of the first three normal vibration modes. Deformation of the interfaces are scaled according to the amplitude of calculated deflection. **d** Brightfield images of a device showing the deflection as a response to excitation at the chosen frequencies. **e** A system is driven with the same signal twice to follow a rectangular trajectory. Rectangle in red dashed lines shows the desired trajectory. **f** A system is driven to follow two circular trajectories with different sizes. Circles

in red dashed lines show the desired trajectories. **g** An actuator module moving along the letters of EPFL. Data points are shown by blue dots and prescribed trajectories are shown by red continuous lines. Scale bar, 20 µm.

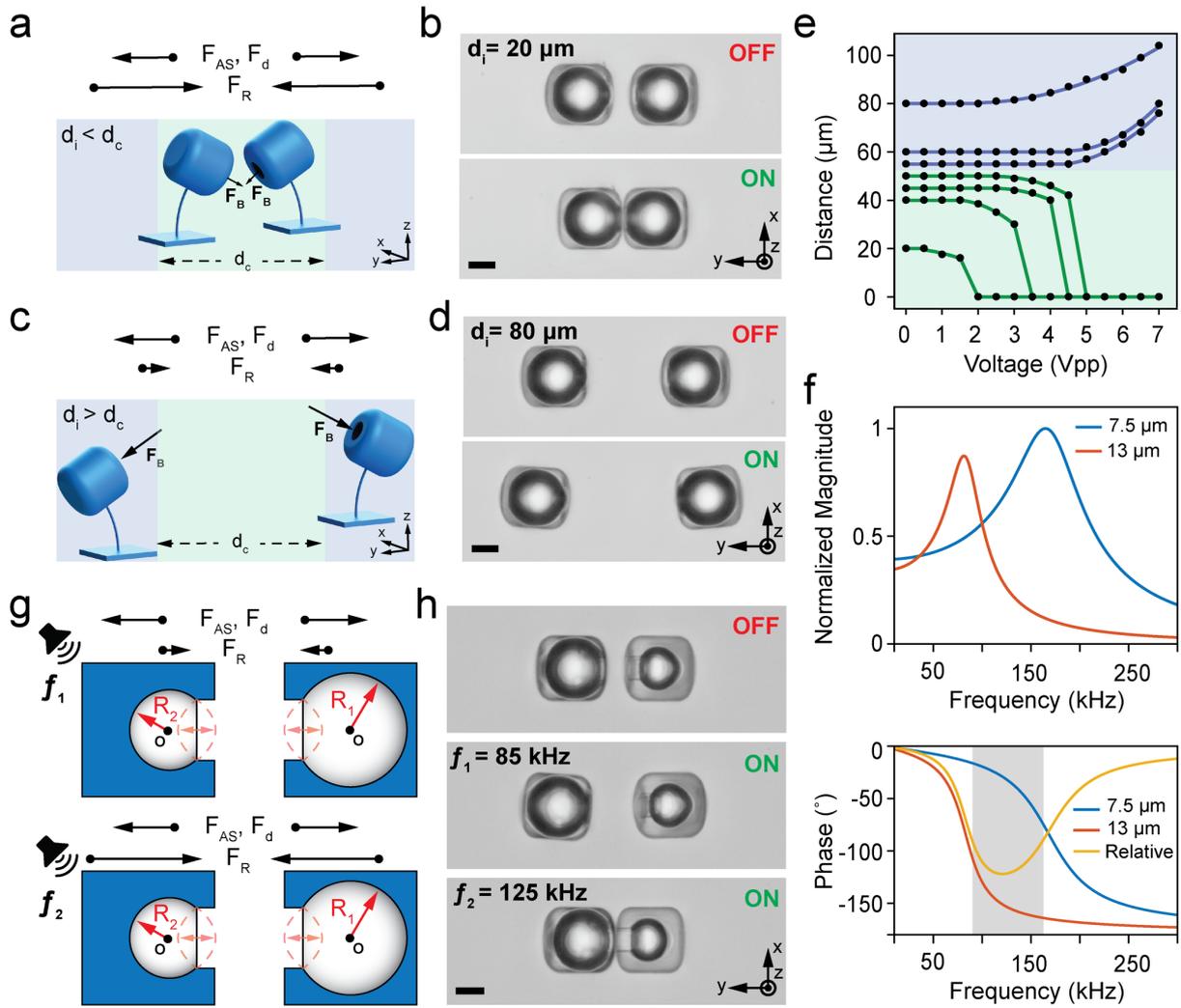

**Fig. 4. Characterization of acoustic radiation forces between two actuator modules. a** Graphical illustration showing the relative magnitude of the three forces acting on the actuator module; drag force ($F_d$), acoustic streaming force ($F_{AS}$), and acoustic radiation force ($F_R$) for the case where the initial distance ($d_i$) between the actuator modules is shorter than the critical distance ($d_c$). In this case, $F_R$ is higher than the sum of the other two forces. **b** Representative microscope images from an experiment where the actuator modules are attracted to each other and, as a result, the beams bend towards each other. **c** Graphical illustration showing the relative magnitude of the three forces acting on the actuator module for the case where $d_i$ is longer than $d_c$. In this case, $F_{AS}$ and $F_d$ dominate the radiation force. **d** Microscope images from an experiment where the actuator modules are moving away from each other and, as a result, the beams bend in opposite directions. **e** The equilibrium distance between acoustically actuated beams as a function of input voltage. The beams are printed with different spacing. In all the experiments, the frequency of the sound wave is tuned to 125 kHz. **f** The theoretical bode plot. The blue and orange curves in the lower panel show the phase response of the small and big bubbles,

respectively. The yellow curve depicts the phase difference. In the gray region, the phase difference results in a repulsive force while in the white regions the interbubble forces are attractive. **g** Schematics showing the frequency-dependent motion of actuator modules with different cavity sizes. The in-phase and out-of-phase vibrations of the entrapped bubble determine the magnitude of the radiation forces. **h** Representative brightfield images showing the frequency-dependent deformation of coupled beams. Scale bars, 20 µm.

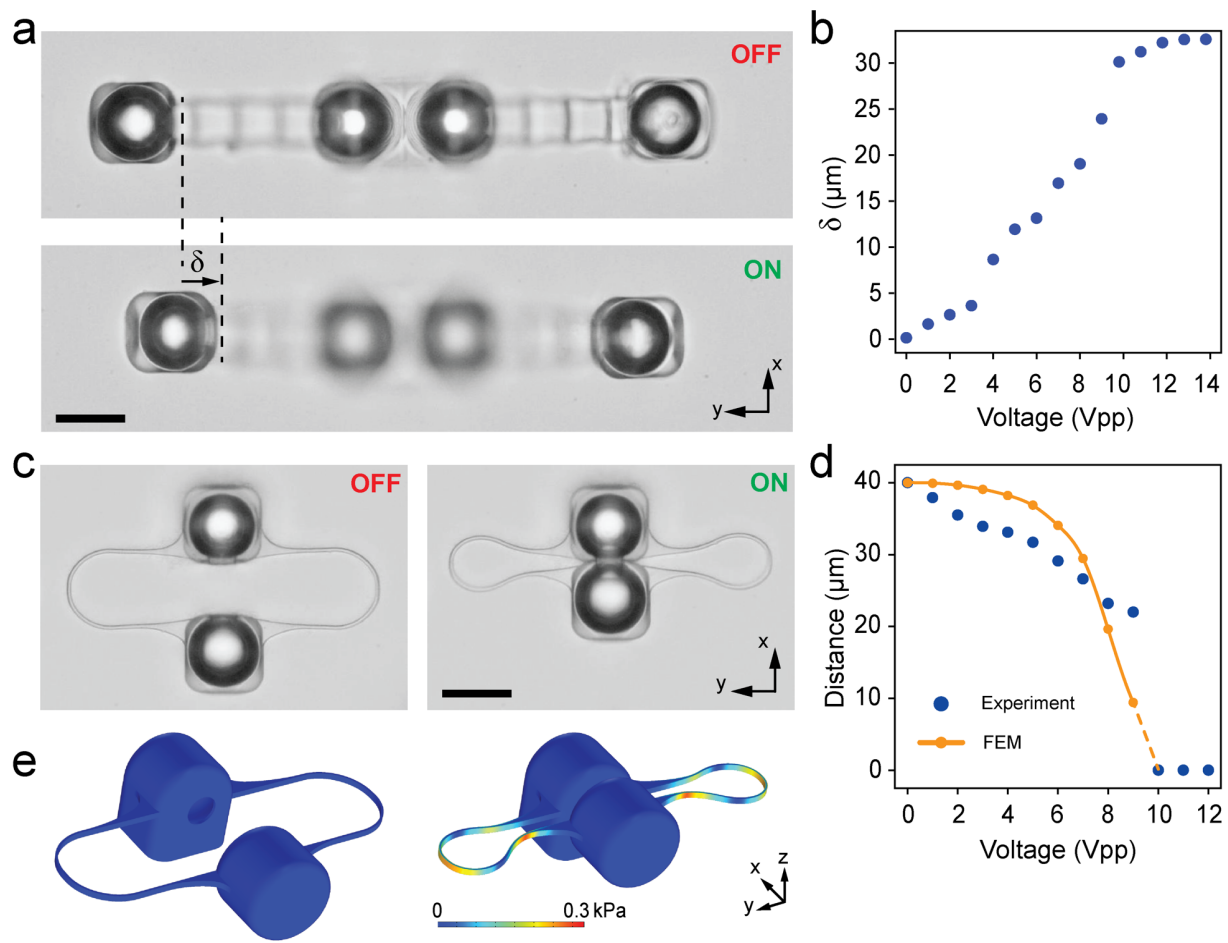

**Fig. 5. Mechanical programming of soft micromachines. a** Out-of-axis deformation of a soft robotic gripper. Two pairs of identical actuator modules are printed on both arms. The displacement, δ, is used to measure the bending of the arms. **b** δ increases monotonically with increasing voltage until a threshold at which the deformation ceases. The radiation forces generated by the entrapped bubbles are insufficient to bend the structure further. The actuator modules are constrained in their motion so that they do not snap with each other. **c** A linear microactuator constructed with a flexible spring mechanism. The top actuator module is fixed to the ground while bottom one is free to move. The motion of the bottom actuator module is constrained in x-axis by the spring mechanism. **d** The distance between the actuator modules gradually decreases with input voltage until the entrapped bubbles get close to each other. Increasing the voltage further abruptly snaps the modules. Simulation results closely match the experimental data. **e** A 3D FEM model of the machine showing the stresses acting on the mechanism. The same model is used to calculate the stiffness of the spring. Scale bars, 50 μm.

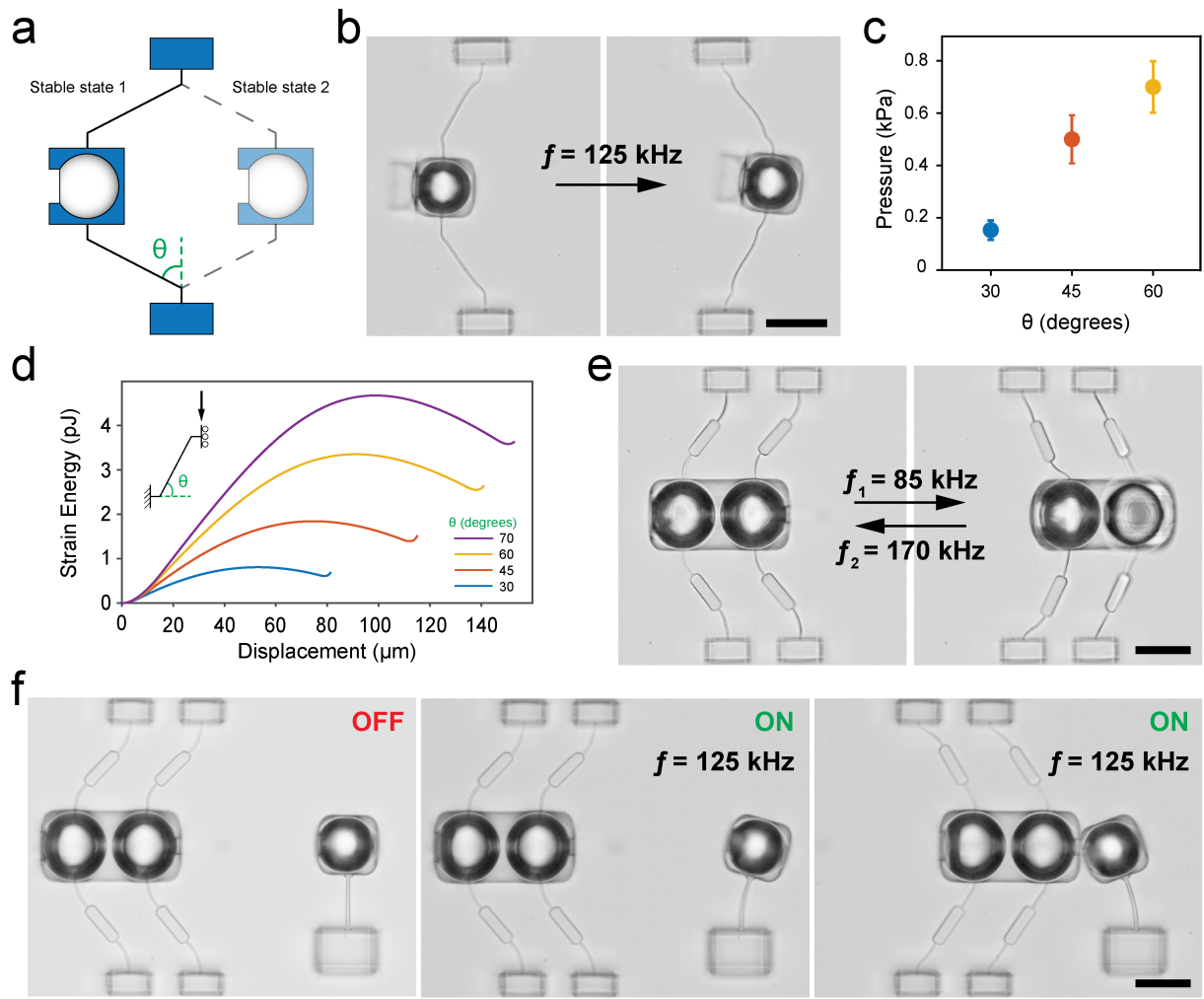

**Fig. 6. Mechanical reprogramming of soft micromachines through elastic instabilities. a** The design of a bistable mechanism with snap through. The inclination angle, $\theta$, is modulated to construct mechanisms with varying energy barriers. **b** Representative microscope images showing the switching of a bistable mechanism from one stable state to another upon acoustic excitation. The mechanism stays indefinitely in both stable states unless it is excited to switch. **c** The acoustic pressure required to pass the energy barrier and switch the mechanism for prototypes with varying $\theta$. **d** The energy landscape of bistable mechanism with different $\theta$. **e** Reversible actuation of the bistable mechanisms. Two actuator modules with identical cavity size and varying orifice sizes are connected on the same unit for frequency selective actuation in opposite directions. **f** A bistable mechanism was used as a control module. A third separate actuator module deformed a cantilever beam in the clockwise direction when the control module is kept at one stable state. Switching the control module to the left completely changes the force balance on the third actuator module. The radiation forces generated between the control module and the third actuator module reverses the motion and deforms the beam in the counter-clockwise direction.

Scale bars, 50 μm.

# Supplementary Information

## 3D printed acoustically programmable soft microactuators

Murat Kaynak[1†], Amit Dolev[1†], and Mahmut Selman Sakar[1*]

[1]Institute of Mechanical Engineering and Institute of Bioengineering, Ecole Polytechnique Fédérale de Lausanne, CH-1015 Lausanne, Switzerland

[†] These authors equally contributed to this work.

**\*Corresponding author:**

Mahmut Selman Sakar, PhD

Institute of Mechanical Engineering, Ecole Polytechnique Fédérale de Lausanne, MED3 2916 Station 9, CH-1015 Lausanne, Switzerland.

Tel: +41 21 693 1095

Email: selman.sakar@epfl.ch

## Supplementary Figures

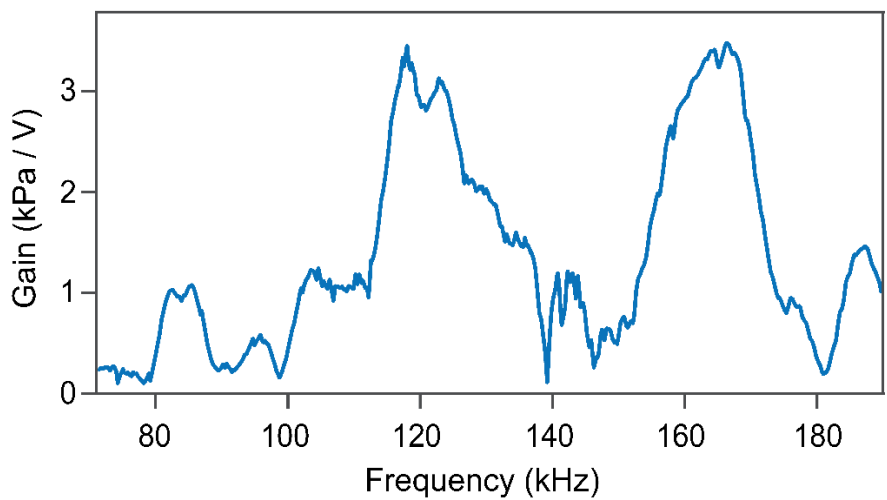

**Figure S1**. Measured transfer function (TF) produced pressure as a result of input voltage. The gain of the TF, which is the pressure produced by the transducer vs. input voltage is plotted as a function of the frequency. The TF captures the dynamics of the transducer, glass slide, and peripheral electronics. Large TF values indicate optimal excitation frequencies.

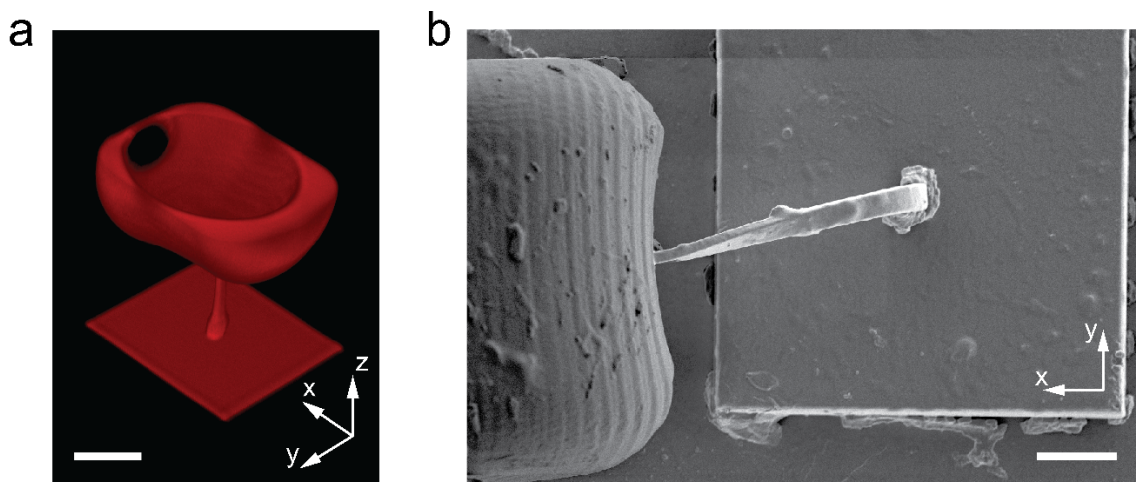

**Figure S2**. Scanning electron and confocal microscopy images of vertically printed actuator module. **a** The vertically printed actuator module was visualized in water using confocal imaging. The top part of the module is not shown to reveal details of the inner cavity and aperture. Scale bar, 20 µm. **b** The size of the beam was confirmed with SEM. Scale bar, 10 µm.

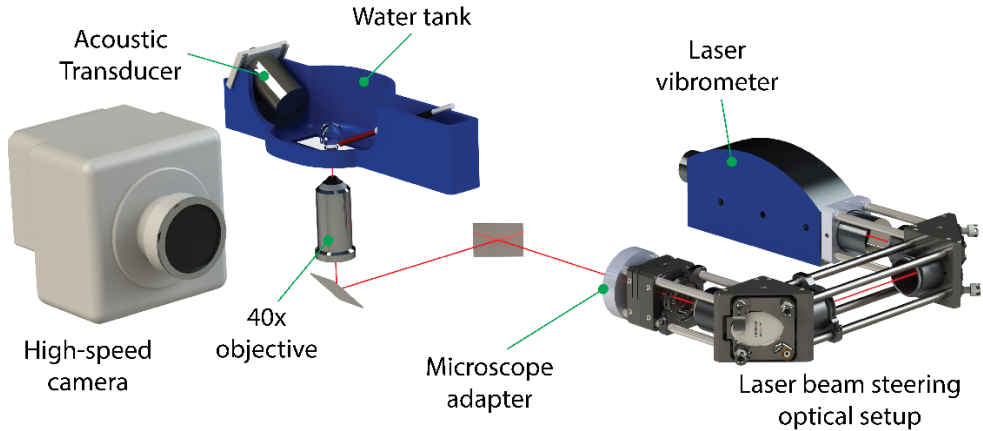

**Figure S3**. Illustration of the experimental setup to measure the bubble deflection. The laser beam emitted by the vibrometer passes through two 45° adjustable mirrors that align and stir the beam. The optical setup is directly connected to an inverted microscope (not shown). The beam passes through two additional 45° mirrors inside the microscope before reaching the sample. The second internal mirror directs the beam towards the microscope objective, incoming light is focused on the sample, and the reflected light is collected by the vibrometer. High-speed imaging is performed using bright-field imaging. The water tank holds an acoustic transducer, a hydrophone, and a glass slide that holds the microfabricated devices.

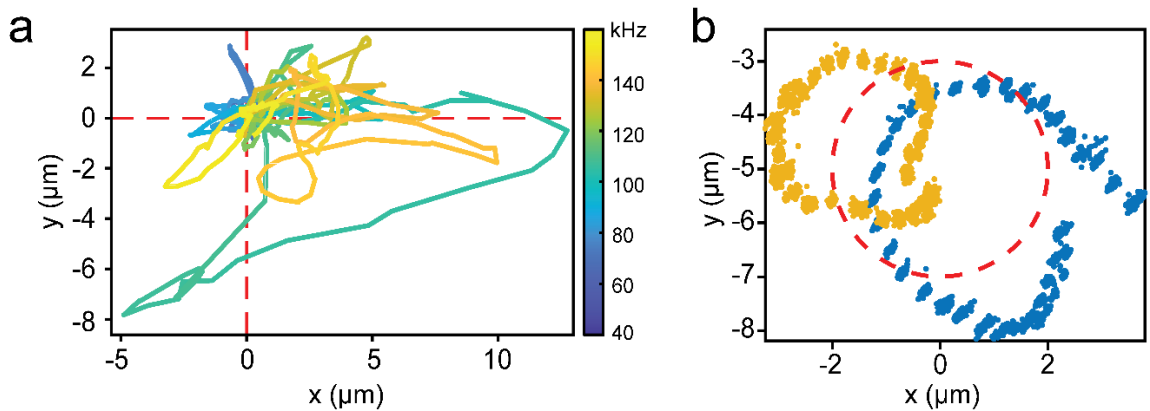

**Figure S4**. **a** The recorded planar position of the device during the frequency sweep from 40 kHz to 160 kHz. **b** The motion of two different devices excited by the same signal. The calibration parameters were extracted from one device (blue) and used for the control of the second device (yellow).

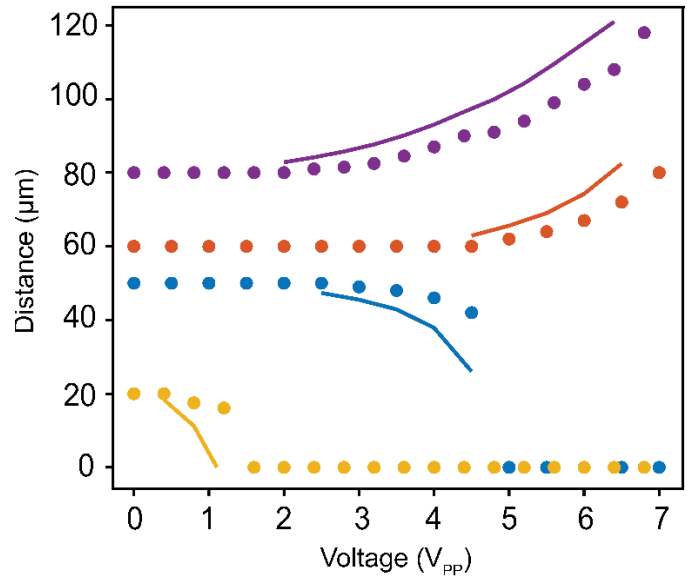

**Figure S5.** The equilibrium distance between acoustically actuated beams as a function of the input voltage for four representative experiments. Measured data is shown by markers, and fitted data is shown by continuous lines. According to the model (Eq. (3) in the manuscript), only the transition region can be fitted.

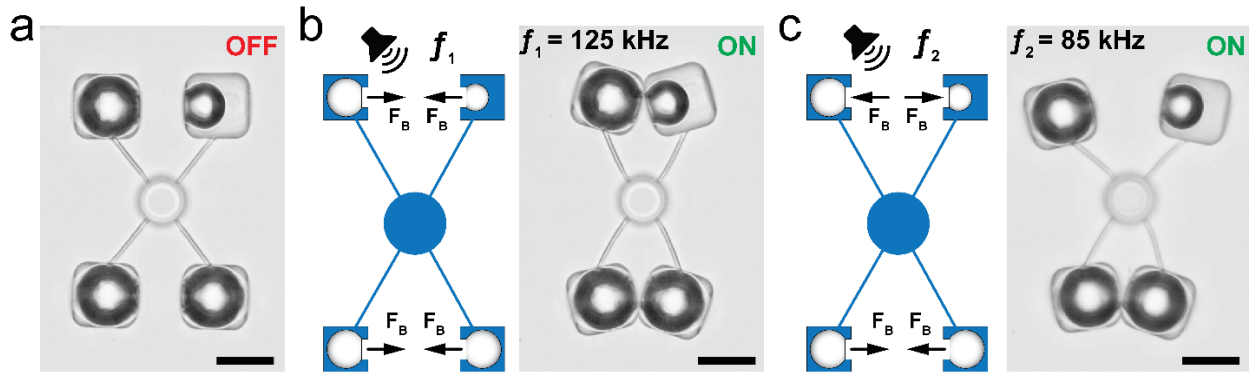

**Figure S6.** Mechanical programming of soft micromachines. **a** Microscope image of a flextensional mechanism with two pairs of actuated beams. Actuator modules on the bottom mechanism are identical while the actuator modules on the top mechanism have different cavity sizes. **b** At 125 kHz, both actuator pairs attract each other. As a result, all arms close. **c** At 85 kHz, the actuator modules located on the top mechanism move away from each other due to the out-of-phase vibrations of the differently sized entrapped bubbles. This motion leads to the opening of the mechanism. The bubbles entrapped by the bottom actuator modules are excited the same way at all frequencies, thus, always move towards each other.

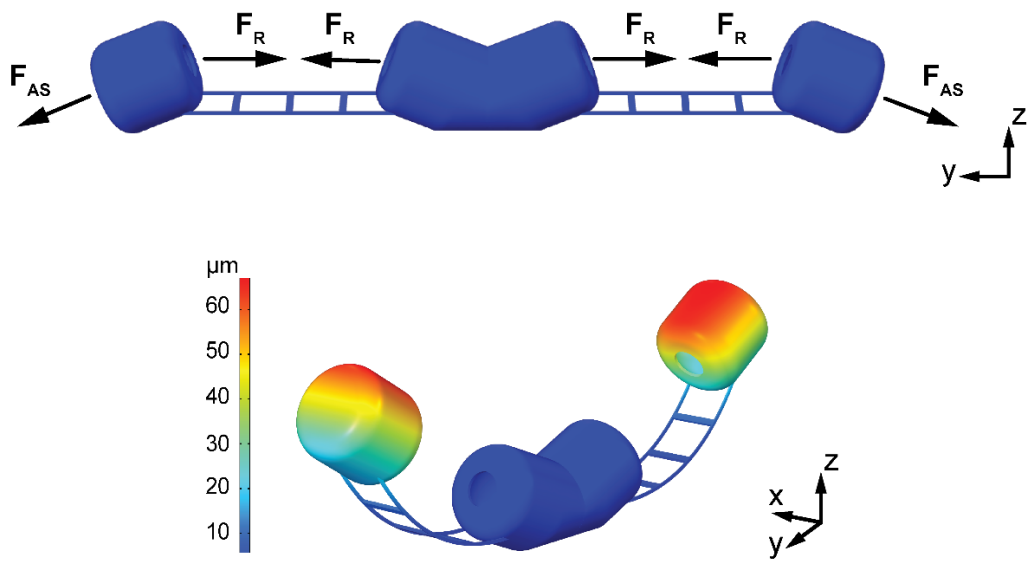

**Figure S7**. **a** Free body diagram and simulation results of an FEM model of the truss mechanism. The input to the model is the experimentally recorded beam deflection. The model calculates $F_R$ by taking experimental characterization of $F_{AS}$, assuming that $F_{AS}$ does not change with beam deformation and decreasing inter-bubble distance. **b** 3D view of the device which is fixed in the middle.

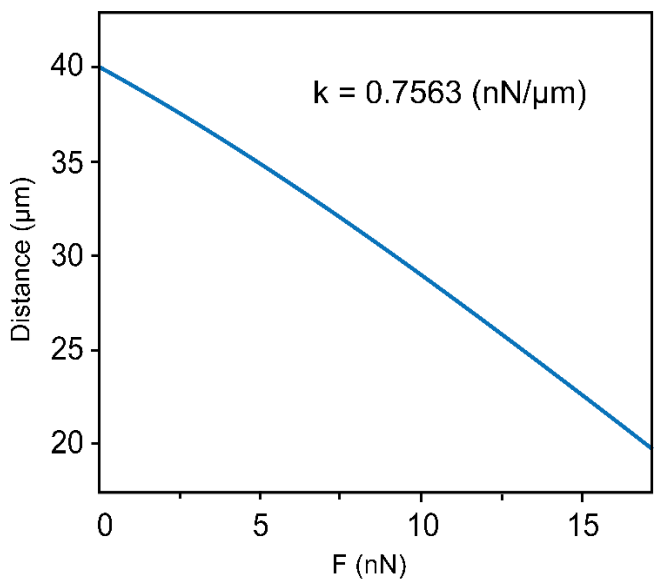

**Figure S8**. The distance between the bubble actuators versus the total bubble-bubble interaction force, computed by the FEM model. The provided stiffness was computed as the mean values of the gradient.

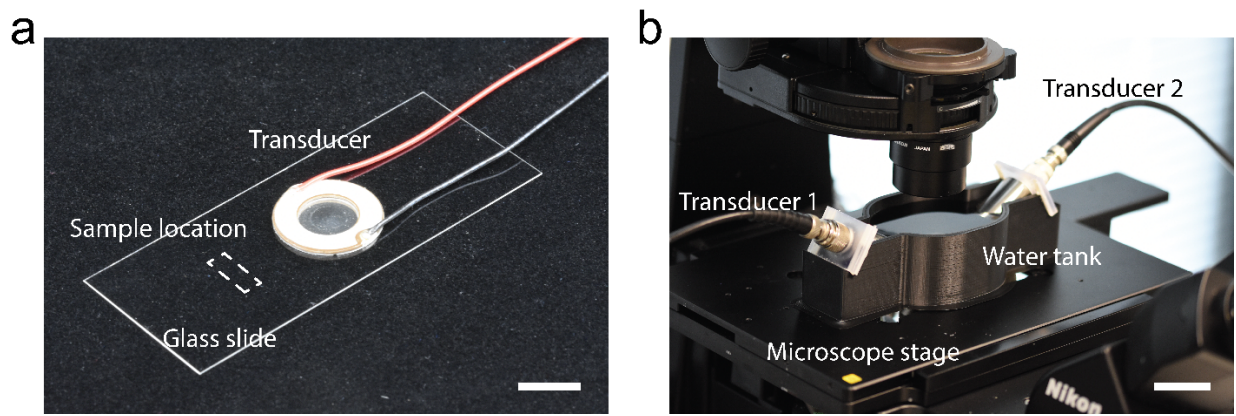

**Figure S9**. Images of the experimental platforms used in this work. **a** A piezoelectric transducer is bonded on a glass slide. The actuator modules are fabricated within the dashed area. Scale bar, 1 cm. **b** Two water immersion ultrasound transducers are submerged into a custom-made water tank. Scale bar, 5 cm.

**Table S1**. Actuator module design parameters

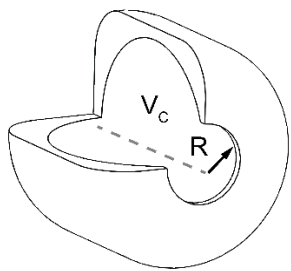

|  | Theoretical Natural Frequency, $f_T$, (kHz) | Excitation Frequency, $f_E$, (kHz) | Bubble volume, $V_c$, (µm³) | Orifice radius, R (µm) |
|---|---|---|---|---|
| Design 1 | 122 | 125 | $6.51 \times 10^4$ | 10 |
| Design 2 | 84.9 | 85 | $1.12 \times 10^5$ | 13 |
| Design 3 | 171.8 | 170 | $1.13 \times 10^5$ | 7.75 |

**Table S2**. The force required to switch the bistable mechanism at different $\theta$.

| $\theta$, ° \ $F_T$, nN | Experimental | Simulation |
|---|---|---|
| 30 | 1.09 | 32.53 |
| 45 | 8.95 | 40.5 |
| 60 | 16.92 | 47.76 |

# Supplementary Notes

## Supplementary Note 1. Experimental Methods

**3D printing of structures.** All reagents were purchased from Sigma Aldrich and used as received, unless otherwise stated. Structures were printed using a 3D laser lithography system (Nanoscribe Photonic Professional GT+) equipped with an oil immersion objective (63x/1.4 DIC M27, Zeiss). The ink was prepared based on a previous publication[1]. Briefly, the polymers (trimethylolpropane ethoxylate triacrylate (TPETA)) and photoinitiator (Irgacure 819) were mixed at a ratio of 98:2 (w/w), and sonicated for 3 hours. A fluorescent die (Rhodamine B) was added to the solution at a final concentration of 0.5% (w/w) for the samples imaged by a confocal microscope. A computer-aided design (CAD) software (Solidworks 2020, Dessault Systèmes) was used to sketch the machines and another software (Describe, Nanoscribe GmbH) was used for file conversion. 3D models were sliced 0.3 µm vertically and 0.2 horizontally, and the laser power and the scan speed were set to 40 mW and 2 mm/s, respectively. The structures were printed in oil immersion mode on a cover glass (No.1, Cat. No. 470819, Brand). To promote surface adhesion, the cover glasses were first activated with a 5-minute plasma treatment (PDC-32G, Harrick Plasma), which was followed by silanization using a solution of (3-(trimethoxylsil) propyl methacrylate, acetic acid, and isopropyl alcohol at a ratio of 3:30:1000 for 1 hour. The printing solution was spin-coated at 1000 rpm for 15 seconds (WS-650-23, Laurell Technologies Corporation) and dried on a hotplate (PZ 28-2 ET, Harry Gestigkeit GmbH) at 100°C for 5 minutes. After printing, the devices were developed in a solution of (Methyl isobutyl ketone (MIBK) and isopropyl alcohol (IPA) at a ratio of 1:1 (v/v)) for 1 hour. The structures were gently rinsed with IPA and air dried.

**Experimental Platform.** The actuators were powered either by a contact (SMMOD15F120, Steminc INC.) or a water immersion (GPS100-D19, Ultran Group) transducer (Fig. S9). The piezo transducer was glued on the cover glass, approximately 5 mm away from the machines, using an epoxy glue (G14250, Devcon). To reduce friction during experiments, the samples were treated with a surfactant (3% Pluoronic F-127 in DI water) for 10 minutes. To increase surface tension and improve bubble stability and lifetime[2,3], phosphate buffered saline (25X PBS, ScyTek Laboratories Inc.) was used as the working solution. A function generator (AFG-2225, GW Instek) that was connected to an amplifier (HVA200 THORLABS) powered the transducers.

**Pressure and vibration measurements**. The pressure around the samples was measured using a needle hydrophone (RP. ACOUSTIC, PVDF hydrophone type l). The output signal was amplified (RP. ACOUSTIC, HVA-10m-60-F) before it was sampled (pico Technology, PicoScope 5243D). The vibration measurements were performed using a laser Doppler vibrometer (Polytec CLV-2534). The laser beam was steered using two 45°-angled mirrors into the objective of an inverted microscope (Nikon, Ti-2) using custom-built adapters. A 40X water immersion objective (CFI Apo NIR, Nikon) was used to reduce the mismatch in the refractive indices, thus providing high enough optical signal for the proper operation of the laser vibrometer. To confirm that the recorded signals are indeed coming from the air-water interface, we performed measurements on several randomly chosen locations from the printed structure and the glass slide by placing reflective stickers.

**Microscopy and imaging.** A high-speed camera (VEO640L, Phantom) connected to an inverted microscope (Ti2, Nikon) with a 20x objective (Nikon) was utilized for image and video acquisition. Videos and imaged were postprocessed with Fiji (National Institute of Health). Data were plotted using a commercially available graphing and data analysis software (Origin 2021, OriginLab Corporation). Flow profile of the streaming was visualized using a 1 μm fluorescent beads (Cat # 17154-10, Polysciences). A scanning electron microscope (Merlin, Zeiss) was utilized to capture high resolution images. Fluorescence images were acquired using an inverted laser scanning confocal microscope (Zeiss LSM 700) equipped with a 20X objective.

**Finite element modelling.** The commercial FEM software COMSOL Multiphysics 5.6 was used to simulate the deformation of structures at the equilibrium state. Large deformations of solid bodies were simulated using the structural mechanics module, where various constraints and loads were applied.

**Statistical analysis.** Results are presented as mean ± standard deviation. Statistical analysis was performed using Origin 2021 (OriginLab Corporation). Data were collected using at least three different trials per device and the number of devices tested (n) is at least 6 for each condition.

## Supplementary Note 2. Computing the beam's stiffness using a linear model

We are interested in deriving the linear stiffness of the beam shown in the figure below assuming linear material and linear geometry.

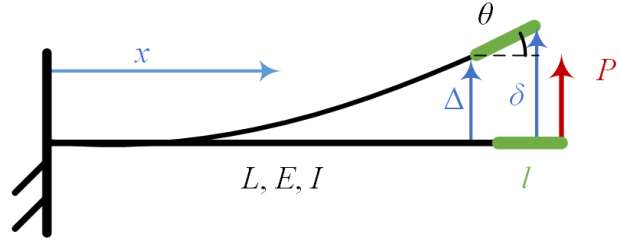

**Figure S10**. Schematics of the free body diagram to derive linear stiffness of the beam.

The black compartment of the beam whose length is $L$ is flexible with young's modulus $E$, and second moment of inertia of the cross-section $I$, while the green part whose length is $l$ is rigid with the same cross section. The beam is subjected to a force acting at its end ($x=L+l$), $P$. To solve only the flexible compartment, we can apply at its end ($x=L$) the force $P$ and a moment equal to $lP$. Therefore, we solve the following linear beam deflection curve:

$$u(x) = a_3 x^3 + a_2 x^2 + a_1 x + a_0. \tag{2.1}$$

The boundary conditions are as follows, at $x = 0$, the beam is clamped:

$$u(0) = u_x(0) = 0, \tag{2.2}$$

and at its end where $x = L$, a force and a moment are applied:

$$u_{xx}(L) = \frac{Pl}{EI}, \quad u_{xxx} = -\frac{P}{EI}. \tag{2.3}$$

The solution yields the following deflection curve

$$u(x) = \frac{P}{6EI}\left[3(l+L) - x\right]x^2, \quad 0 \le x \le L. \tag{2.4}$$

Therefore, the linear stiffness and slope at the end of the beam are:

$$k = \frac{6EI}{L^2(3l+2L)}, \quad u_x(L) = \tan(\theta) = \frac{L(2l+L)P}{2EI} \tag{2.5}$$

The total deflection at the end of the rigid beam is

$$\delta = \Delta + l\sin(\theta) = \Delta + l\sin\left[\arctan\left(\frac{L(2l+L)P}{2EI}\right)\right] = \Delta + \frac{lL(2l+L)P}{EI\sqrt{4 + \frac{L^2(2l+L)^2 P^2}{EI^2}}} \quad (2.6)$$

Since $\Delta = P/k$.

$$\delta = P\left[\frac{1}{k} + \frac{lL(2l+L)}{EI\sqrt{4 + \frac{L^2(2l+L)^2 P^2}{(EI)^2}}}\right] = \frac{LP}{6EI}\left[L(3l+2L) + \frac{6l(2l+L)}{\sqrt{4 + \frac{L^2(2l+L)^2 P^2}{E^2I^2}}}\right] \quad (2.7)$$

Notice that $\delta$ is not a linear function of $P$.

## Supplementary Note 3. Fitting the acoustic radiation force model

For the out of plain mechanism, we fitted a simplified model of the acoustic radiation force, $F_R = \beta V^2 d^{-2}$. The fit is good ($R^2 = 0.85$), however a better model can be fitted ($R^2 = 0.99$) as shown in **Error! Reference source not found.**, $F_R = \beta V d^{-2}$. The better fit of the second model can be attributed to several factors, to understand them we begin by analyzing the secondary radiation force acting between two identical spherical bubbles:

$$F_R = -4\pi\rho\omega^2 \frac{R_0^2 \varepsilon^2}{d^2} \qquad (3.1)$$

The force depends on the fluid density $\rho$, excitation frequency $\omega$, oscillation amplitude $\varepsilon$, and the geometry through the distance between the bubbles $d$, and their radius at rest $R_0$. Furthermore, this force is correct only for spherical bubbles and it is different for other geometries. Therefore, it is reasonable to assume that the better fit obtained for the second case does not indicate a physical relation. It is possible that $\beta$ is a function of the geometry, and as the mechanism deforms it should change as well. Also, in the simulation we did not consider the streaming generated by the anchored bubble whose projection on the moving one changes as the mechanism deforms. To summarize, because not all physical phenomena were modeled a wrong model that matches the data can be fitted, but not justified physically.

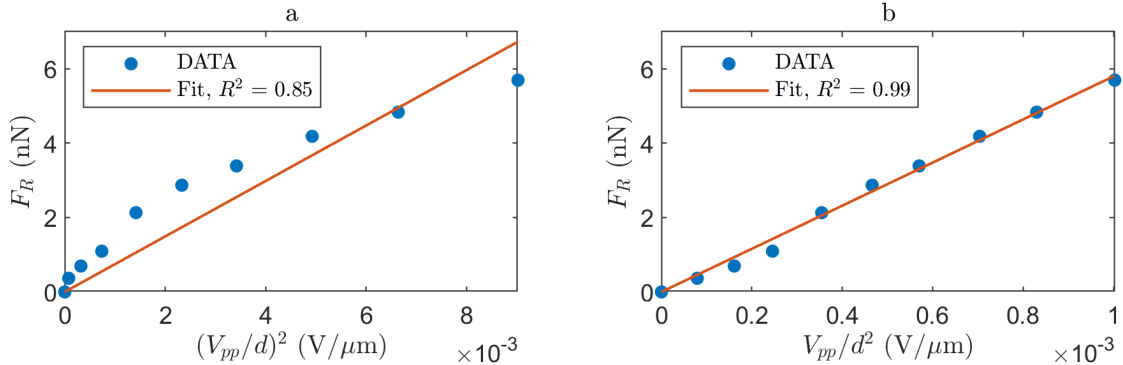

**Figure S11**. Fit of the acoustic radiation force using two models. In panel a, the theoretical model is $F_R = \beta V^2 d^{-2}$, and in panel b the model is $F_R = \beta V d^{-2}$. The $R^2$ values are given for both cases.

# Description of Additional Supplementary Files

**File Name:** Movie S1

**Description:** Entrapped bubbles actuated selectively at their natural frequencies of either 85 kHz or 170 kHz, which leads to sequential deformation of the beams.

**File Name:** Movie S2

**Description:** An actuator module with three orifices driving motion around a prescribed circular trajectory.

**File Name:** Movie S3

**Description:** Direction of $F_R$ between the same actuator modules switches with respect to distance. Actuator modules facing each other either attract or repel when $d_i < d_c$ or $d_i > d_c$, respectively.

**File Name:** Movie S4

**Description:** Direction of $F_R$ between actuator modules of same opening size and different volume, switches according to excitation frequencies. The in-phase and out-of-phase vibrations of the entrapped bubble determine direction of the radiation forces.

**File Name:** Movie S5

**Description:** Frequency dependent behavior of a flextensional mechanism. All four arms simultaneously closed at 125 kHz. Then, one couple opened while the other closed at 85 kHz.

**File Name:** Movie S6

**Description:** Out-of-axis deformation of a soft robotic gripper using acoustic radiation force. Upon acoustic excitation, two pairs of identical actuator modules which are printed with an angle, attracted each other and bend the beams out of axis.

**File Name:** Movie S7

**Description:** A linear microactuator constructed with a flexible spring mechanism. The top actuator module is fixed on the ground while the motion of bottom actuator module is free to move in x-axis.

**File Name:** Movie S8

**Description:** Reversible actuation of bistable mechanisms. Two actuator modules with identical cavity size and varying apertures are connected on the same unit for frequency-selective actuation in opposite directions.

**File Name:** Movie S9

**Description:** A bistable mechanism as a control module. A third separate actuator module deformed a cantilever beam in the clockwise direction when the control module is kept at stable state 1. The radiation forces generated between the control module and the third actuator module reverses the motion and deforms the beam in the counter-clockwise direction when the control module is at stable state 2.